# Hierarchical self-assembly of nanoparticles for optical metamaterials

Sergio Gomez-Graña[1], Aurélie Le Beulze[1], Mona Treguer-Delapierre[1], Stéphane Mornet[1], Etienne Duguet[1], Eftychia Grana[2], Eric Cloutet[2], Georges Hadziioannou[2], Jacques Leng[3], Jean-Baptiste Salmon[3], Vasyl G. Kravets[4], Alexander N. Grigorenko[4], Naga A. Peyyety[5], Virginie Ponsinet[5], Philippe Richetti[5], Alexandre Baron[5], Daniel Torrent[5], Philippe Barois[5]*

[1] University of Bordeaux, CNRS, ICMCB, UPR 9048, 33600 Pessac, France.
[2] University of Bordeaux, CNRS, LCPO, UMR 5629, 33615 Pessac, France.
[3] University of Bordeaux, CNRS, Solvay, LOF, UMR 5258, 33608 Pessac, France.
[4] School of Physics and Astronomy, University of Manchester, Manchester, M13 9PL, UK.
[5] University of Bordeaux, CNRS, CRPP, UPR 8641, 33600 Pessac, France.

* e-mail: barois@crpp-bordeaux.cnrs.fr

**Hierarchical self-assembly arranges nanostructures at different length scales. It gradually becomes an effective method of fabricating artificial metamaterials from composite nanostructures tailored for a particular response. Hierarchical self-assembly overcomes shortcomings of "top-down" methods by significantly reducing fabrication time and making it possible to form bulk materials. Here we report an application of hierarchical self-assembly of metal nanoparticles for the creation of the first isotropic optical metamaterial with strong artificial magnetism in blue light. We have used colloidal self-assembly to create artificial "nanomolecules" that generate the desired magnetic response and microfluidic self-assembly to produce a bulk metastructure. We demonstrate that the magnetic response of the final material is accurately described by an isotropic magnetic permeability that satisfies the principle of locality. Our approach unlocks the fabrication of large volumes of composite nanomaterials. Moreover, the spatial disorder inherent to this "bottom-up" method holds the key to solving the non-locality problem. The technique can be readily extended to the future generations of low-loss optical metamaterials made of dielectric nano-blocks to bypass the limitations of optical losses associated with plasmonic resonances in noble metals.**

Many optical metamaterials require structuring on at least two different length scales: one to produce "nanomolecules" that generate the necessary optical response at the sub-wavelength scale, and another to assemble these "nanomolecules" into a macroscopic metastructure exhibiting the desired optical properties. One of the best examples of tiered structuring is given by a magnetic metamaterial where the "molecules" responsible for the artificial magnetic response need to be geometrically complex since the magnetic permeability may be viewed as the long-wavelength limit of spatial dispersion. Hierarchical self-assembly is precisely suited to tackle such situations [1-4]. It is fast, cheap and extremely effective for production of bulk metastructures. Moreover, the structural disorder, which is sometimes cited as a drawback of bottom-up self-assembly methods, appears to be a considerable advantage in this case as it provides natural isotropy. Here we make use of the hierarchical self-assembly strategy to develop a bulk isotropic magnetic material which is accurately depicted by a bulk magnetic permeability. This type of material is actually absent despite intensive research and huge progress in optical metamaterials in recent years. It is fair to say that the ability to control independently the values of the dielectric permittivity ($\varepsilon$) and of the magnetic permeability ($\mu$) of optical materials would open extraordinary applications that have been abundantly discussed in the field of metamaterials. The perfect lens, based on a hypothetical negative index material resulting from simultaneous negative values of $\varepsilon$ and $\mu$ [5], and the invisibility cloak, based on the concept of transformation optics implying the realization of well controlled gradients of $\varepsilon$ and $\mu$ [6] were seminal proposals of the field. Applications of transformation optics are foreseen not just in cloaking, which consists of bending light around an obstacle to make it invisible, but also in light concentrators, for the generation of intense fields for sensing or non-linear optics, or for optical impedance matching, enabling a more efficient collection of light in optical devices [7]. Controlling $\mu$ is notoriously difficult since the magnetic susceptibility of natural materials is negligible in visible light [8]. Consequently, the generation of optical magnetism has been intensively searched in artificial nanocomposites, generally designed so as to develop loops of intense induced currents upon illumination by a light wave [9-14]. While the use of noble metals has unveiled the fundamental limitations due to optical losses associated with plasmonic resonances, the perspective of future all-dielectric optical devices based on Mie resonances [15,16] restores the interest of understanding the light-matter interaction leading to artificial magnetism. The self-assembly method described here is indeed applicable to the next generation of all-dielectric metamaterials.

All the appealing properties of magnetic metamaterials cited above rest upon the implicit assumption that the magnetic permeability $\mu$ maintains in artificial composites its familiar meaning based on the introduction in Maxwell's equations of the auxiliary field $\mathbf{H}=\mathbf{B}/\mu_0\mu$ and $\mathbf{D}=\varepsilon_0\varepsilon\mathbf{E}$ deriving from the alleged structure of the induced current density $\mathbf{J} = \nabla\times\mathbf{M} + \varepsilon_0\varepsilon\partial\mathbf{E}/\partial t$ in which $\mathbf{M}$ identifies with a local density of magnetization. This is unfortunately not true in general since the space and time derivatives of the induced currents cannot be separated. A more rigourous approach consists in using a generalized non local dielectric tensor $\varepsilon_{ij}(\omega,\mathbf{k})$ accounting for the wave vector $\mathbf{k}$ dependence of the electromagnetic response of the composite material – referred to as spatial dispersion – in addition to the ω frequency dispersion [17,18]. In the case of weak spatial dispersion, the frequency and wave vector dependences can be separated in a power-series expansion $\varepsilon_{ij}(\omega,\mathbf{k}) = \varepsilon_{ij}(\omega) + \beta_{ijl}(\omega)k_l + \gamma_{ijlm}(\omega)k_l k_m$. For achiral systems with spatial inversion symmetry, $\beta_{ijl}(\omega) = 0$ and the artificial magnetism is depicted by the fourth rank tensor $\gamma_{ijlm}(\omega)$, making it obvious that a second rank permeability tensor $\mu_{ij}$ will not suffice to describe the optical properties. The number of independent components of a fourth rank tensor actually depends on the symmetry of the composite structure: it ranges from 36 in a triclinic system to 4 in a cubic symmetry and reduces to 2 in an isotropic structure with C∞ continuous spherical symmetry [17]. For a second rank tensor, the number of independent elements is respectively 6, 1 and 1. Consequently, any attempt to describe the artificial optical magnetism by the sole permeability tensor $\mu_{ij}$ amounts to neglecting a set of parameters which is prohibitively large in systems of low symmetry, but which interestingly reduces to a single component in isotropic systems, that thus appear as the best candidates to give some substance to the permeability approximation. Up to now, the observation of strong artificial magnetism at frequencies of visible or near-infrared light has been reported in several types of plasmonic nanostructures fabricated by top-down lithography techniques [10-14]. They are essentially two-dimensional and highly anisotropic. Among them, the most symmetric one is the fishnet structure of tetragonal symmetry [13], for which fourth rank tensors still contain 7 independent parameters, whereas only two are allowed in a second rank permeability tensor. The inadequacy of the permeability tensor to describe the strong artificial magnetism observed in fishnet devices is discussed in [19].

The material described in the present work is built on a different rational approach designed so as to produce a nanostructure with a perfect spherical symmetry in order to check the validity of the magnetic permeability parameter, which ought to be a scalar in this case, under the most favourable symmetry conditions, as discussed above. Our bottom-up strategy consists in a hierarchical self assembly process starting with the multi-step chemical synthesis of spherical nano-resonators followed by the self-assembly of these meta-atoms in a random close packing. The structure of the meta-atoms was first suggested theoretically by Alu,

Salandrino and Engheta [20] and perfected by Simovski and Tretyakov [21] in its isotropic form of raspberry-like magnetic nanoclusters (MNCs, Fig. 1a-b). Each MNC consists of a set of metallic nanoparticles evenly distributed on the surface of a core dielectric sphere. The magnetic response of the MNC is generated by plasmonic currents circulating in the corona of metallic satellites under visible light excitation near the plasma frequency. The observation of a magnetic response was indeed reported in dilute solutions of such MNCs [22-24] or related nano-objects [25]. Here we apply our dedicated hierarchical bottom-up self-assembly approach to fabricate dense three-dimensional samples of sufficient thickness from raspberry-like MNCs. Our artificial materials behave as semi-infinite samples for the reflected wave, which allows a direct analytical retrieval of the effective parameters $\varepsilon$ and $\mu$ from experimental measurements using variable angle spectroscopic ellipsometry. The locality of the isotropic structure is directly demonstrated by the absence of angular dependence of the extracted parameters. The fabricated material exhibits an isotropic effective magnetic permeability $\mu$ ranging from 0.8 to 1.45 at visible wavelengths between 250 and 800 nm.

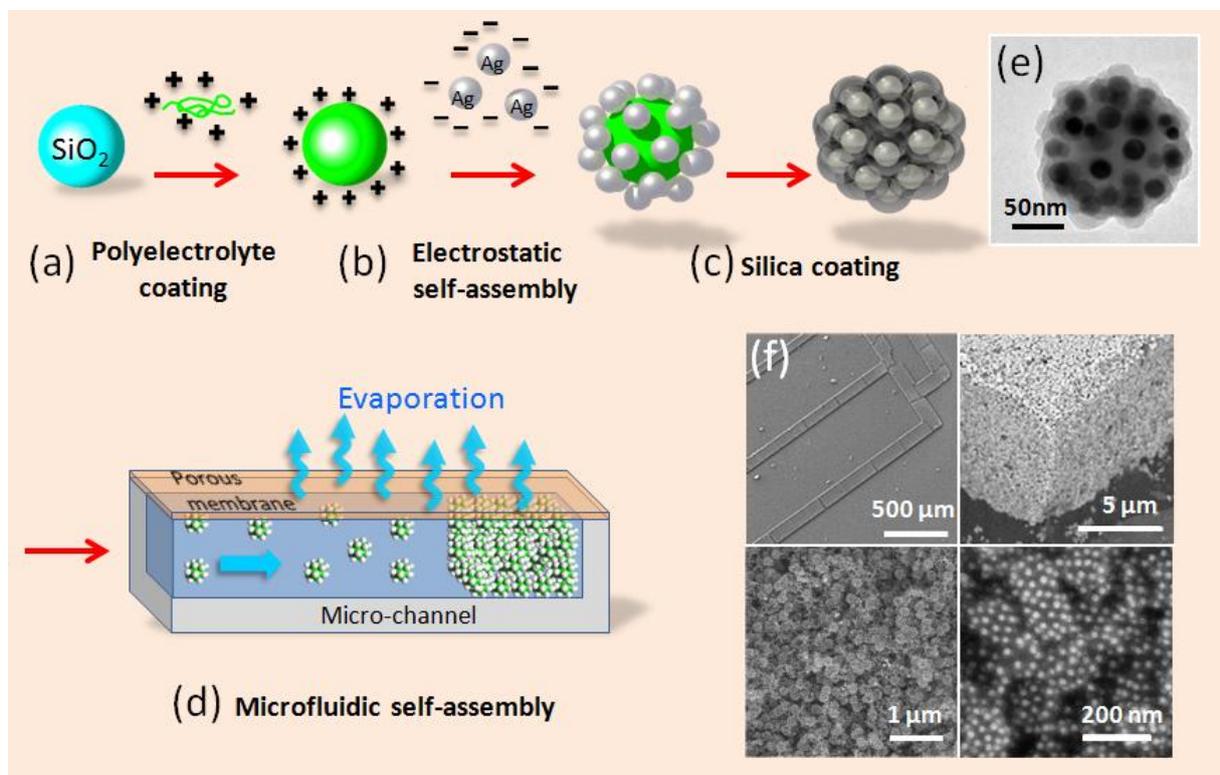

**Figure 1: Fabrication of the self-assembled magnetic metamaterial.** (a-d) Sketch of the hierarchical self-assembly process. Silica nanoparticles coated by a polycation in (a) self-assemble with negatively charged silver nanoparticles in (b) to form raspberry-like meta-molecules similar to the Simovski-Tretyakov model of magnetic nanocluster (MNC). An outer silica layer is grown in (c) to secure the nanostructure. The MNCs are subsequently self-assembled in a microfluidic evaporator (d) to form a bulk metamaterial. (e) Transmission Electron Microscopy micrograph of a real MNC. (f) High-Resolution Scanning Electron Microscopy images of the resulting material at different scales, showing the neat geometrical features of the dense solid made of randomly packed MNCs.

We start the fabrication of the metamaterial with the synthesis of the meta-atoms. Large amounts of MNCs are synthesized in aqueous dispersions by electrostatic self-assembly of positively charged silica cores of diameter 96 ± 3 nm and negatively charged silver satellites of diameter 25 ± 2 nm. The positive charge of the silica core is obtained by layer-by-layer adsorption of a cationic polyelectrolyte onto the silica surface [26]. The negative surface charge of the silver satellites results from the stabilizing polymer agent used for the synthesis. The details of the synthesis are given in supplementary information. A fine tuning of the ratio of concentration of cores and satellites during the self-assembly process yields a narrow size distribution of MNC clusters [24,27]. A thin layer (5 nm) of silica is grown on the surface of the clusters to secure their morphology and improve their colloidal stability. A TEM micrograph of the as-obtained MNCs is shown in Fig. 1b.

Having produced dilute dispersions of MNCs, we formulate the bulk metamaterial in a microfluidic device by slow evaporation of the solvent across a thin polymer membrane (see Fig. 1c, supplementary information and references [28-30] for details). The final 3D-material is a solid phase made of densely packed MNCs fitting the shape of the micro-channel. The volume fraction of the MNCs is estimated to be 60(±5)%, which is consistent with random close packing of spheres. Fig. 1e reveals the excellent definition of the outer shape of the final 3D material, perfectly replicating the micro-channel template with a surface roughness set by the size of the MNCs.

We resort to variable angle spectroscopic ellipsometry to retrieve experimentally the effective electromagnetic parameters of the formulated metamaterial. We measure the complex ellipsometric ratio $\rho = r_p/r_s$ where $r_p$ and $r_s$ are the amplitude reflection coefficients for light polarization parallel (*p*), and perpendicular (*s*) to the plane of incidence (Fig. 2a). The data analysis is based on two key features: (i) the sample is thicker than the absorption length of the material and behaves as a semi-infinite material and (ii) the Fresnel formulae are written as functions of the impedances $Z = (\mu/\varepsilon)^{1/2}$ to account for variable permeability. A simple relationship between the measured ellipsometric ratio $\rho$ and the complex material parameters $\varepsilon$ and $\mu$ extracted from the Fresnel formulae is as follows

$$\left(\frac{1-\rho}{1+\rho}\right)^2 \frac{\sin^4\theta}{\cos^2\theta} = -A\sin^2\theta + B \qquad \text{with} \qquad A = \left(\frac{\mu-\varepsilon}{1-\mu\varepsilon}\right)^2, B = \varepsilon\mu\left(\frac{\mu-\varepsilon}{1-\mu\varepsilon}\right)^2, \qquad (1)$$

where $\theta$ is the angle of incidence in the ambient air medium of refractive index 1.

Equation 1 has a linear form $Y=AX+B$ that enables the extraction of the unknown coefficients $A$ and $B$ for each wavelength from a linear regression of the experimental quantities $X$ and $Y$ measured on a set of five different angles separated by 5° from 50 to 70° (Fig. 2b). Inverting the pair of quadratic equations for $A$ and $B$ yields a set of 4 solutions ($\varepsilon$, $\mu$) among which the physical ones are selected so as to satisfy the physical constraints that the imaginary part $k$ of the refractive index and the real part $Z'$ of the optical impedance $Z$ should be positive. For example, the complex refractive index can be found directly as $N^2=\varepsilon\mu=B/A$. The physical solutions are displayed for two different positions in the sample in Fig. 2c-f. Those data points for which the Pearson correlation coefficient [31] of the linear regression is higher than 0.95 are specified by circles to highlight the good quality of the linear fit to equation (1) across the whole spectrum. More details on the extraction of the permittivity and permeability from ellipsometry measurements are given in the supplementary information.

Figures 2c-f show the spectral variations of the effective parameters $\varepsilon$ and $\mu$ in the 250-800 nm range. Both parameters exhibit a resonant behavior centered at $\lambda \cong 420$ nm. The magnetic permeability deviates notably from the natural value $\mu = 1+0i$ with a real part $\mu'$ ranging from 0.8 to 1.45 and a negative imaginary part $\mu''$. These variations reveal a real part of the magnetic susceptibility $\chi_m = \mu - 1$ reaching a negative peak value of approximately −0.2 at $\lambda = 460$ nm, three orders of magnitude stronger than the record natural diamagnetic susceptibility of graphite [32], and a positive peak value of +0.45 at $\lambda = 400$ nm, opposite to the classical negative sign of diamagnetism.

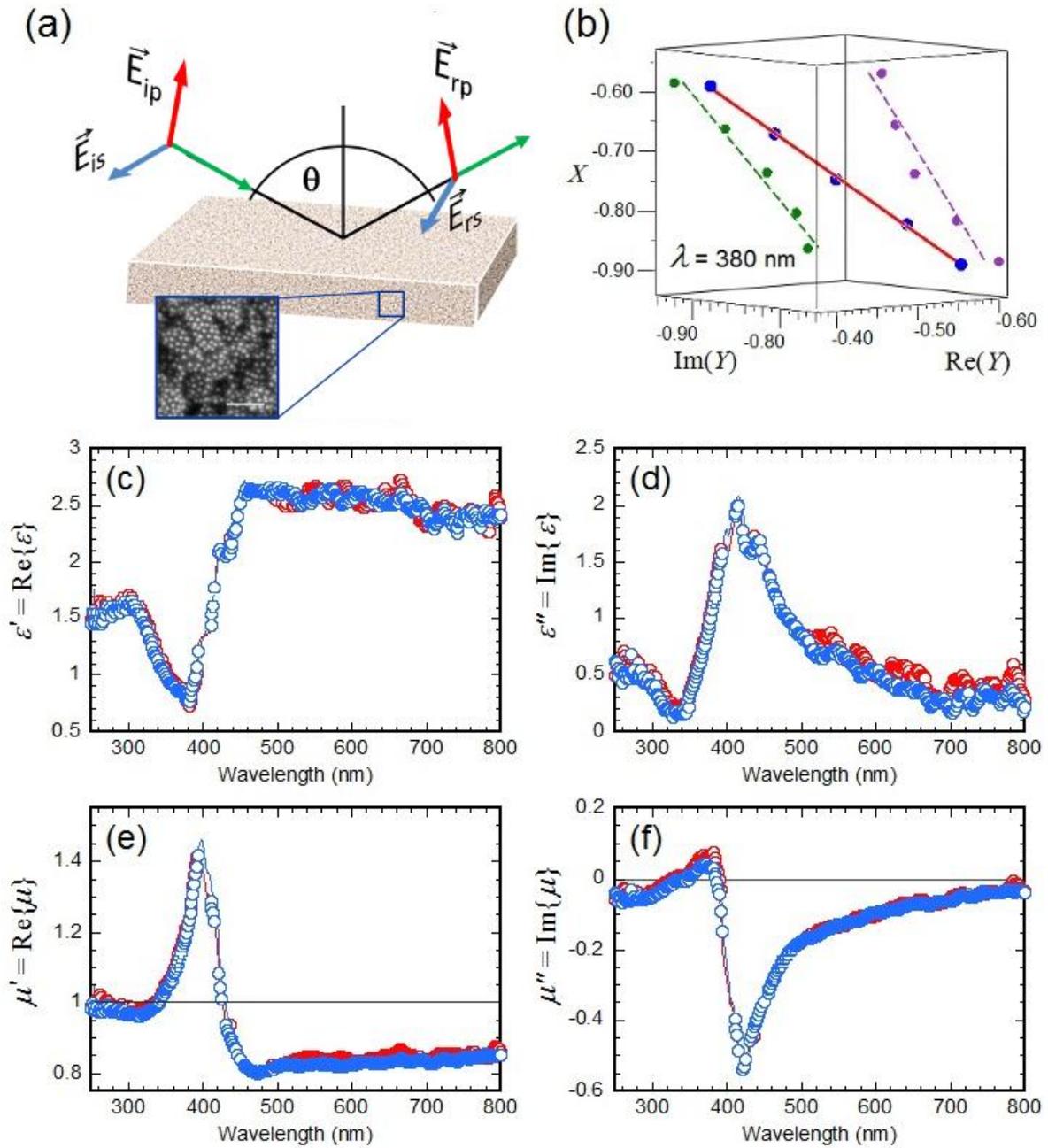

**Figure 2: Electromagnetic parameters of the self-assembled metamaterial.** (a) Geometry of the ellipsometric experiment. (b) Illustration of the linear relationship $Y=AX+B$ (see text) at a particular wavelength $\lambda = 380$ nm (solid line). Dashed lines show the projections of the linear form onto the planes $(X, \text{Re}(Y))$ in purple and $(X, \text{Im}(Y))$ in green. (c,d) and (e,f) show the real and imaginary part of the permittivity $\varepsilon$ and of the permeability $\mu$ respectively. Two different colours correspond to data collected by spectroscopic ellipsometry at two different positions in the sample. Solid lines display all data whereas the circles denote those points for which the Pearson correlation coefficient of the linear regression (equation (1)) is higher than 0.95.

Finding negative values of the imaginary part $\mu''$ of the magnetic permeability may look surprising. It is not unique though, since negative values of $\mu''$ were already reported in numerous theoretical and numerical studies of metamaterials [33,34]. The dielectric and magnetic responses of the MNCs are coupled plasmonic processes generated by the same unique driving force, namely the electric field of the light wave. Physical laws impose that the sum of these two processes must be dissipative. It is indeed the case since the imaginary part $k$ of the refractive index and the real part $Z'$ of the impedance are positive, see the discussion below. The goodness of the linear fit to equation (1) throughout the wavelength range is evidenced in Fig. 2c-f by the large number of circles denoting the data with a Pearson correlation coefficient larger than 0.95. It demonstrates that the parameters $(\varepsilon,\mu)$ do not depend significantly on the direction $\theta$ of the incident wave vector **K** in the experimental range. The major result of this ellipsometric study is therefore the demonstration that the metamaterial exhibits giant effective isotropic diamagnetism without any measurable effect of spatial dispersion.

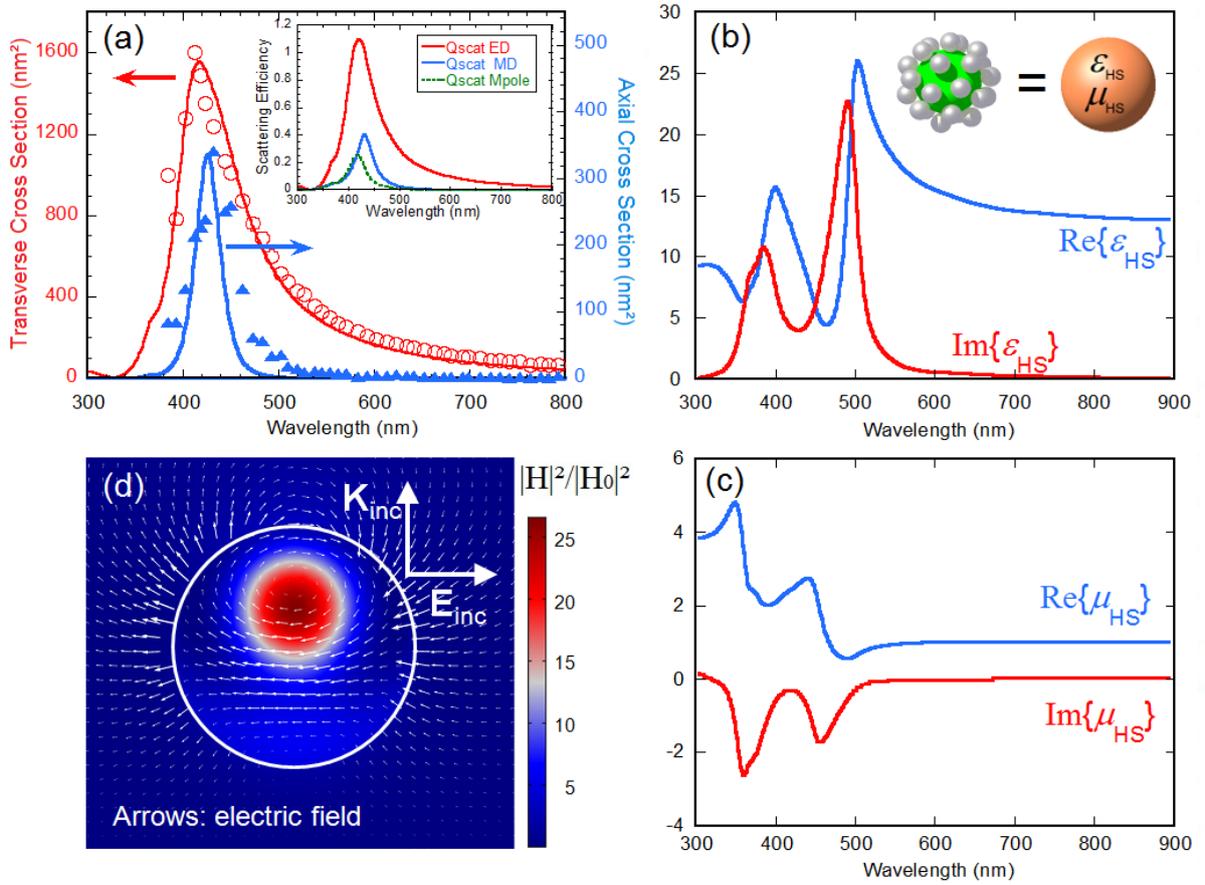

**Figure 3: Electromagnetic response of a single magnetic nanocluster.** (a) Experimental transverse (circles) and axial (triangles) scattering cross-sections measured at 90° scattering angle on a dispersion of MNCs. Solid lines are T-Matrix numerical simulations of a MNC constructed along the model of figure 1a-b with 22 silver satellites. Inset shows scattering efficiencies of the dipolar and quadrupolar modes. (b-c) Spectral variations of the real and imaginary parts of the permittivity and the permeability of the homogeneous sphere (HS) equivalent to the MNC. (d) Map of the squared magnetic field (color scale) and electric field (arrows) within the HS showing that the effective particle acts as a magnetic dipole at the magnetic resonance $\lambda$ = 420 nm.

Next, we test the order of magnitude of the experimentally retrieved optical constants by comparing them to those retrieved from full-wave numerical simulations of the self-assembled metamaterial. Following a method described in reference [24], the scattering of the MNC is computed by a T-Matrix code [35]. This method accurately reproduces the electric and magnetic dipolar scattering measured experimentally (Fig. 3a). The numerical expansion yields the electric and magnetic dipolar scattering coefficients of a single MNC $a_1$ and $b_1$ [35,36] and enables the modeling of the self-assembled material by the finite-element method. A direct simulation, however, requires a three-dimensional assembly model of the MNC, which is prohibitive in computational time owing to the complexity of the meta-atom. The physics of the problem can still be captured while greatly reducing the computational load by replacing the MNC (made of 22 silver spheres plus one silica core) with a single fictitious homogeneous sphere (HS) with effective parameters ($\varepsilon_{HS}$, $\mu_{HS}$) chosen so as to produce exactly the same dipolar scattering as the MNC [19]. Mie theory provides exact expressions for the scattering coefficients $a_1$ and $b_1$ of the homogeneous sphere [36], which we use to determine ($\varepsilon_{HS}$, $\mu_{HS}$) by solving numerically the two equations $a_1$(MNC) = $a_1$(HS) and $b_1$(MNC) = $b_1$(HS) (see supplementary information). The resulting exact numerical solutions $\varepsilon_{HS}(\lambda)$ and $\mu_{HS}(\lambda)$ are displayed in Fig. 3b,c. Both material parameters show a doubly resonant behavior associated with strong spectral variations of the effective permeability and permittivity. The field profile in Fig. 3d shows the scattered magnetic field $|\mathbf{H}|^2$ along with the spatial distribution of the scattered electric field $\mathbf{E}$ at $\lambda$ = 430 nm, corresponding to the magnetic dipole resonance wavelength. These maps provide a clear signature of the magnetic dipolar resonance. They evidence the circulating electric currents which generate the effective diamagnetic response of the assembled metamaterial.

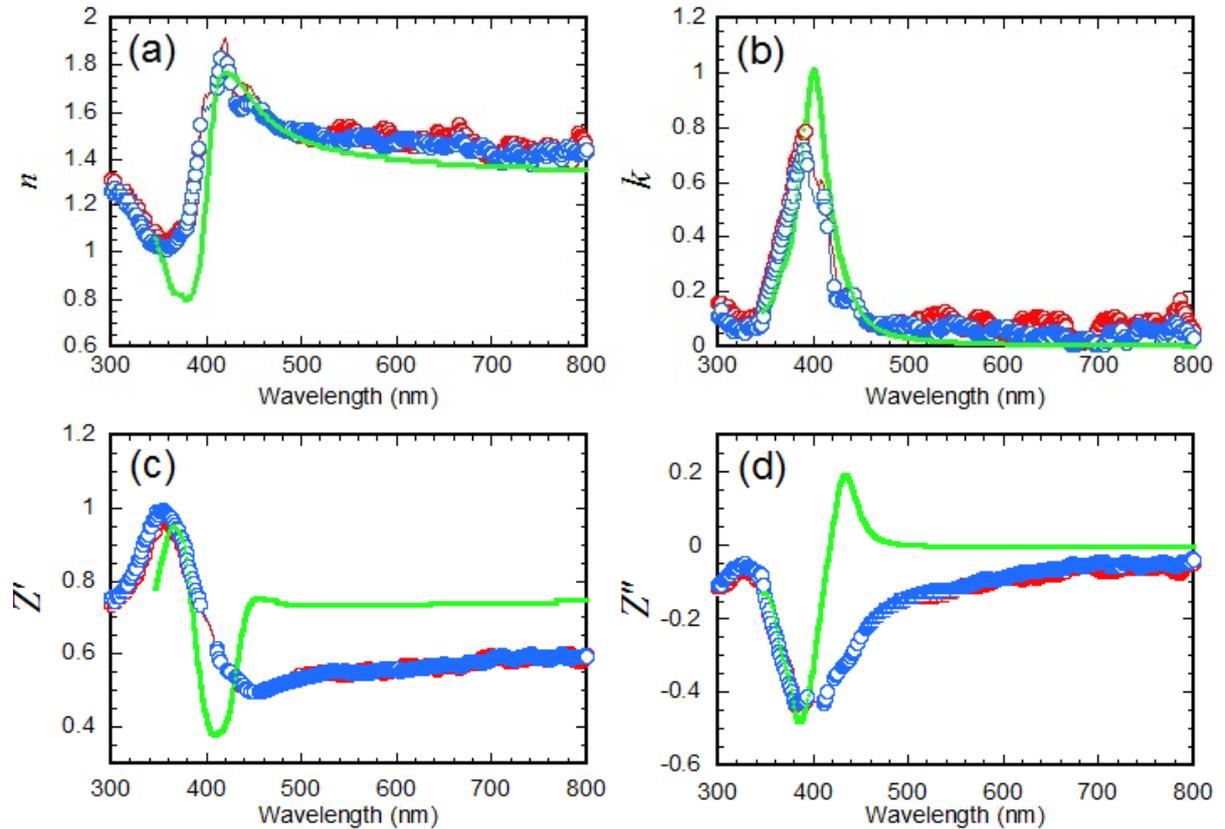

**Figure 4: Comparison with numerical simulations.** Real and imaginary part of the refractive index (a, b) and of the optical impedance (c, d) of the metamaterial. Symbols and solid lines in blue and red correspond to the experimental data shown in figure 2, whereas the green line is a finite element numerical simulation with COMSOL.

Finally, we simulate a three dimensional arrangement of HS's on a cubic lattice to compute the response of a sample of assembled MNCs. The period $a$ = 107 nm is set equal to the distance between closest neighbours in the compact arrangement. We use the finite element based commercial software COMSOL Multiphysics to compute the reflectance and transmittance coefficients under normal incidence. Following the classical retrieval method described by Smith *et al.* [37], we compute the spectral variations of the effective refractive index $N=n+ik$ and effective impedance $Z=Z'+iZ''$ of the metamaterial from the scattering matrix. Figure 4 compares the retrieved parameters to the experimental values extracted by variable angle spectroscopic ellipsometry. The agreement is qualitatively good: the numerical and experimental values of the four functions *n, k, Z'* and *Z''* display very similar spectral variations across the whole visible range. Moreover, the quantitative agreement is remarkably good for the refractive index, see Fig. 4a-b. To the best of our knowledge, this is the first time that the values of the effective parameters of a magnetic metamaterial obtained with the normal angle retrieval method [37] and scattering simulations [24] are confirmed by variable angle measurements (see the discussion in ref. [38]). This further strengthens our claim that the bulk self-assembled metamaterials fabricated in our work indeed satisfy the criterion of effective locality.

In conclusion, through a multidisciplinary approach combining nano-chemistry, colloidal physics, spectroscopic ellipsometry, light scattering and numerical simulations, we demonstrate that isotropic artificial optical magnetism can be produced in a bulk metamaterial at visible light frequencies with no significant effect of spatial dispersion. Our bottom-up approach is based on a hierarchical self-assembly strategy whereby metallic and dielectric nanoparticles are first assembled to form isotropic magnetic nanoclusters of sub-wavelength size and are subsequently concentrated in micro-channels to form a dense three-dimensional material. The spatial disorder of the final material, inherent in self-assembly techniques, warrants the continuous spherical symmetry of the nanostructure, which we identify as a *conditio sine qua non* to minimize the non-locality phenomenon. We take advantage of the strong absorption associated with the plasmonic resonances to study our materials in the optically simple regime of a semi infinite medium. Losses are however not welcome and dielectric nanoresonators will undoubtedly serve as better candidates than plasmonic systems for future applications. This work should for example stimulate research for the synthesis of silicon nanospheres of well controlled size, which are not yet available in large amounts [15,16]. Whatever the nature of the spherical meta-atom, this work paves the way to the large scale production of magnetic metamaterials of arbitrary shape and size, as required for optical hardware and the manipulation of the flow of light by transformation optics [7].

**Acknowledgements:**
We thank Mirela Malekovic for the synthesis of the silver satellites and Pr. Constantin Simovski for enlightening discussions about optical magnetism. EG, EC and GH acknowledge financial support from the Industrial Chair (Arkema/ANR) within the grant agreement number. AC-2013-365.
This work was supported by the LabEx AMADEus (ANR-10-LABX-42) in the framework of IdEx Bordeaux (ANR-10-IDEX-03-02), France.

# Supplementary information

Note: additional references cited the supplementary section are denoted by [Sx].

## 1 – Synthesis of meta-atoms: Experimental section

Chemicals: Silver nitrate ($AgNO_3$, 99.999%), sodium hydrosulfide (NaHS), L-arginine (≥ 98%), tetraethyl orthosilicate (TEOS; ≥ 90%), sodium citrate dihydrate (≥ 99%), cyclohexane (> 99.7%), poly-(vinyl-pyrrolidone) (PVP; Mw~55,000) and Acetone ($CH_3COCH_3$, ≥99.5%) were purchased from Sigma-Aldrich. Iodomethane ($CH_3I$, 99%) was purchased from Acros Organics and N,N-Dimethylformamide (DMF, anhydrous 99.8% packaged under Argon in resealable bottle) from Alfa Aesar. Ethylene glycol (EG, 99%), ammonium hydroxide ($NH_4OH$, 28-30 wt.%) were obtained from J.T. Baker. Absolute ethanol was purchased from Atlantic Labo. Poly(1-vinyl-1,2,4-triazole) (PVTri, MW=21,260 g/mol calculated through conversion from $^1$H-NMR) was prepared through Reversible Addition Fragmentation Chain Transfer (RAFT) polymerization at 60°C in DMF [S1].

Synthesis:

*Cationic polymer electrolyte:* Poly(1-vinyl-4-methyl-1,2,4-triazolium iodide) ($PVMeTri^+I^-$) was synthesized through quaternization of the corresponding neutral polymer [S2]. 400 mg of PVTri (4.2 mmol) is refluxed under inert atmosphere with a large excess (1 mL, 16 mmol) of $CH_3I$ in DMF (10% w/v). Final product is recovered by precipitation in 250 mL acetone and then is dried at 40°C for 20h (726 mg, yield 72.7%, 82.7% quaternized product calculated from $^1$H-NMR data, see Fig. S1). $^1$H NMR (400.1 MHz, NS=16, DMSO-d6) $\delta$ (ppm): 10.4-8.8 (2H, cationic polymer ring), 8.7-7.5 (2H, neutral polymer ring), 4.8-4.1 (1H, cationic polyelectrolyte chain), 4.1-3.5 (1H, neutral polymer chain / 3H, $CH_3$ cationic polymer), 2.5-1.8 (2H, polymer chain both neutral and cationic).

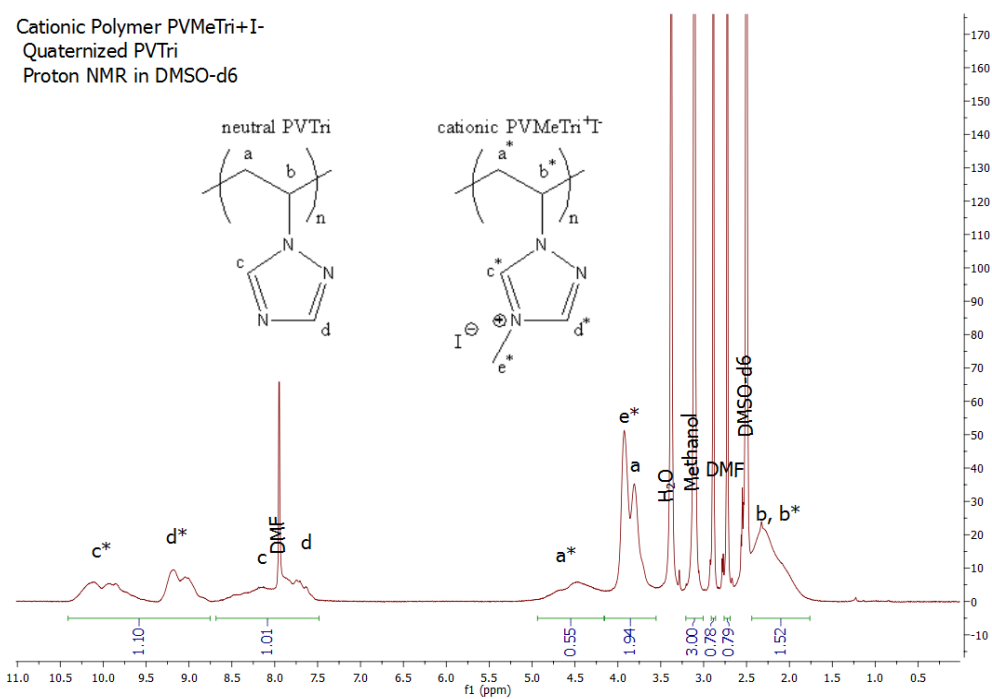

*Figure S1:* $^1$H-NMR spectrum for the cationic polymer electrolyte. NMR measurements were carried out at 298 K on a Bruker Avance spectrometer operating at 400.1 MHz with NS=16. The relaxation time was set at 2s and DMSO-d6 was used as an internal reference ($\delta$=2.50 ppm).

*Silver nanoparticles:* Silver nanospheres were synthetized following to the procedure of Massé et al. [S3] In two different vials, 16.9 mg of NaHS were dissolved in 10 mL of EG and 900 mg of PVP in 30 mL of EG. Both solutions were left to age for 4 h under magnetic stirring. After 1h 30 min, 120 mL of EG was introduced in a rounded flask, being equipped with a reflux condenser and the temperature was increased to 150°C. An Argon flux was introduced in the rounded flask with the EG for the last 30 minutes in order to create an inert atmosphere. After 4 hours of aging, 70 μL of the NaHS solution and 30 mL of the PVP solution were added. Immediately after, 10 mL of EG solution containing 240 mg of silver nitrate was added to the rounded flask. After 10 min, the reaction was quenched introducing the rounded-beaker in an ice-bath and adding 50 mL of cold absolute ethanol. Finally, the silver nanoparticles were washed 3 times by using a Millipore Stirred Ultrafiltration Cell with a regenerated cellulose membrane (100 kDa). Then the nanoparticles were concentrated to ~10 mL by using the ultrafiltration technique.

*Silica nanoparticles:* 345 mL of an aqueous solution of L-arginine (7.5 mM) and 22.5 mL of cyclohexane were introduced in a 500-mL rounded beaker equipped with a magnetic stirrer and the resulting biphasic liquid was heated at 60°C. Then, 37.5 mL of TEOS were added to the top organic phase. The stirring speed was modified in order to form an interface with a constant area between both phases. After 24h, the cyclohexane was removed in a rotary evaporapor at 50°C. A further growth of the silica nanoparticles until 100nm was performed using the Stöber method: 50 mL of ethanol, 5 mL of NH4OH (1 M), and 5 mL of the aqueous suspension of silica seeds were mixed in a 150-mL round-beaker. A pre-determined volume of TEOS diluted in 20 mL of absolute ethanol was added at a rate of 1 mL/h by an automatic syringe pump.

*Layer-by-Layer (LbL) adsorption on silica nanospheres:* The LbL technique was applied according to a previously reported protocol [24]. Firstly, the silica nanoparticles were washed 3 times by centrifugation at 9000 g for 10 min, the pellet was redispersed in ultrapure water. 10 mL of the silica nanoparticles were added to 20 mL of a solution of cationic polymer (2 mg/mL). The adsorption of the polymer onto the silica nanospheres was allowed to proceed for 30 min in a roller mixer. The excess of cationic polymer was removed by centrifugation at 6,000 g for 10 min, the pellet was redispersed in 10 mL of water.

*Synthesis of Raspberry-like nanoclusters:* Typically, 990 μL of water was mixed with 10 μL of the polyelectrolyte-modified silica beads, adding the solution drop-wise in 9 mL of washed silver nanoparticles solution. The mixture was allowed to stir overnight at room temperature and protected from light exposure to avoid the silver oxidation. The excess of metallic nanoparticles was eliminated by centrifugation at 1500 g during 10 min, the pellet with the nanoclusters was redispersed in 1 mL of water. A silica shell was further added in order to keep the raspberry shape. Such silica shell was done by adding 10 mL of ethanol, 250 μL of ammonium hydroxide and 1 μL of TEOS to the solution. The reaction was kept overnight and a centrifugation was performed to remove the free silica nuclei in excess. A further concentration of the sample was performed to obtain a concentrated sample (40 μL at volume fraction of order 1%) for the microfluidics assemblies.

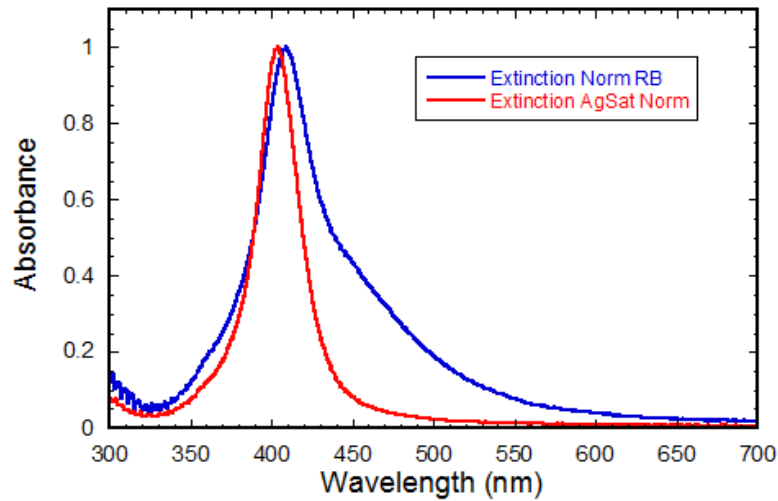

*Figure S2:* *Extinction spectra of a suspension of Ag satellites (red) and MNCs (blue).*

## 2 – Microfluidic assembly:

Microfluidic evaporation cells were created using standard soft photolithography techniques. A master template defining the micro-channels was first made with a photocurable resist polymer. A polydimethylsiloxane (PDMS) thin membrane (e ≈ 30 μm) was prepared by spin coating a PDMS solution onto the master template followed by curing at 60°C. At the same time, a thick PDMS stamp was done by curing the PDMS in a Petri dish at 60°C. Using plasma activation, the PDMS stamp was pasted on the PDMS membrane at one end of the micro-channels. The membrane was then peeled off from the template and punched to create an opening for the reservoir. The PDMS microfluidic chip was deposited on a microscope slide, allowing the evaporation of the solvent only through the membrane.

The MNCs dispersion was introduced in the reservoir, and the channels filled up with the MNCs solution. Due to the evaporation of the water through the PDMS membrane the MNCs accumulate at the dead end of the channel, hence forming the 3D material replicating the channel shape and size. When the reservoir is completely empty, the PDMS membrane is peeled off the glass substrate, leaving the metamaterial on the glass slide.

## 3 – Imaging:

Transmission electron microscopy (TEM) images were obtained with a JEOL JEM 1400 transmission electron microscope operating at an acceleration voltage of 120 kV. Scanning electron microscopy (SEM) images of assembled 3D metamaterial have been obtained using a JEOL JSM-6700F FEG scanning electron microscope operating at an acceleration voltage of 15.0 kV for secondary-electron imaging (SEI).

## 4 – Light scattering experiment:

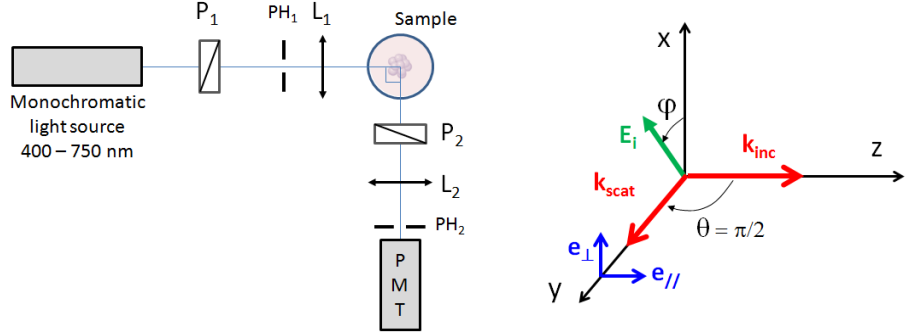

*Figure S3:* *Schematic view of the light scattering setup (left) and scattering geometry (right). The rotation of the input polarizer $P_1$ controls the incident polarization angle $\varphi$. The output polarizer $P_2$ enables to collect the scattering along the transverse ($e_\perp$) and axial ($e_{//}$) directions. PH1 and PH2 are pinholes, L1 and L2 focusing lenses.*

A supercontinuum white source (SuperK EXB-6 with SuperK Varia monochromator from NKT Photonics) was used to deliver a monochromatic light beam. The adjustable bandwidth was set to 6 nm. The linear polarization of the incident light was set by a Glan-Taylor polarizer (Thorlabs GL5-A) mounted on a rotation stage to rotate the incident polarization by an angle $\varphi$. The residual polarization of the source produced a weak variation of the irradiance of order +/- 4% upon rotation $\varphi$ which was corrected in data analysis. The scattered light was collected with a photomultiplier tube (PMT Hamamatsu H10682-01) coupled with a photon counting unit (Hamamatsu C8855-01) at a fixed scattering angle $\theta = 90°$. The polarization of the scattered light was analyzed by a second Glan-Taylor polarizer (Thorlabs GL10-A) set perpendicular or parallel to the scattering plane to collect the $I_{\perp S}$ and $I_{//S}$ signal respectively. The intensities measured along the two output polarizations read:

$$I_{\perp S}(\varphi) = I_0(\lambda)\frac{N_{MNC}}{k^2}|S_1(\theta=90°)|^2 \, \delta\Omega \, T(\lambda) \, g(\lambda,\delta\Omega) \times \sin^2\varphi = A_\perp(\lambda)\sin^2\varphi \quad (S1a)$$

$$I_{//S}(\varphi) = I_0(\lambda)\frac{N_{MNC}}{k^2}|S_2(\theta=90°)|^2 \, \delta\Omega \, T(\lambda) g(\lambda,\delta\Omega) \times \cos^2\varphi = A_{//}(\lambda)\cos^2\varphi \quad (S1b)$$

in which $I_0(\lambda)$ is the spectral irradiance of the incident beam, $N_{MNC}$ is the number of MNC scatterers in the scattering volume, $\delta\Omega$ is the solid angle of the detection window. $g(\lambda,\delta\Omega)$ is an unknown function that accounts for the spectral sensitivity of the detector and optical transmission or reflection of all optical elements. The spectral transmission $T(\lambda)$ account for the absorbance of the suspension.

The $I_{\perp S}$ and $I_{//S}$ signals are shown in inset of Fig. S4 versus the direction $\varphi$ of the incident polarization at $\lambda = 444$ nm. They are fitted for each wavelength to $\cos^2\varphi$ and $\sin^2\varphi$ functions respectively. The ratio of axial to transverse scattering $A_{//}(\lambda)/A_\perp(\lambda)$ measuring unambiguously the ratio of the MD and EQ contributions to the ED scattering is shown in Fig. S4. It exhibits a sharp maximum of 28% at 450 nm, far higher than any reported value so far [23,24].

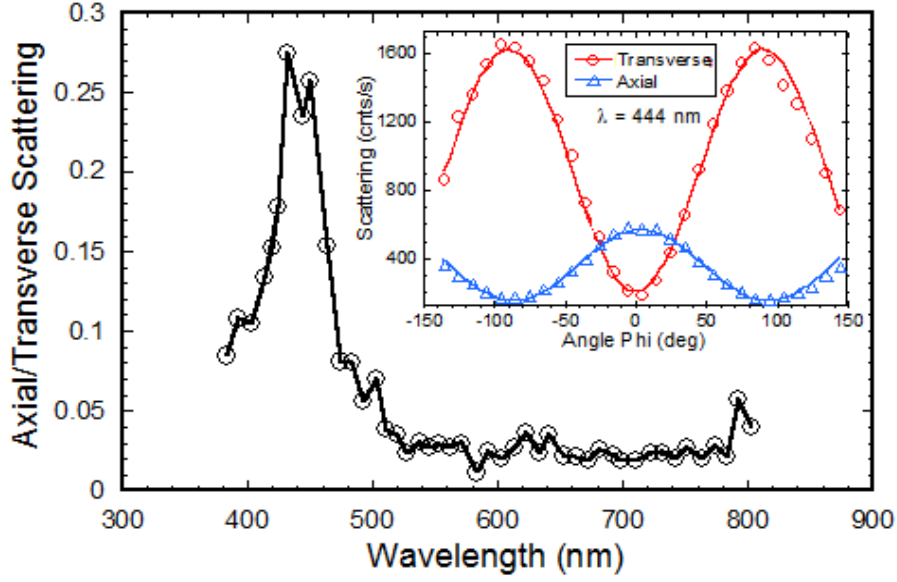

*Figure S4:* Ratio of axial to transverse scattering $A_H/A_V$ vs wavelength. Inset shows transverse (red) and axial (blue) scattering measured at $\lambda=444$ nm. Solid lines are fits to $\sin^2$ and $\cos^2$ functions respectively.

The MD+EQ and ED contributions can be extracted separately by normalizing the data with a reference dispersion of calibrated silica nanoparticles ($D_{SiO2} = 96$ nm) in water. The scattered intensity from the reference sample is similar to equation (S1a) by replacing $|S_1(\theta = 90°)|^2/k^2$ by $\sigma_{ref}(\theta = \varphi = 90°)$ the differential scattering cross section of a silica particle in water at scattering angles $\theta = \varphi = 90°$. The MNC and reference signals were collected in the same experimental conditions so that the quantities $I_0$, $\delta\Omega$ and $g(\lambda,\delta\Omega)$ are the same for MNCs and reference. Dividing equations (S1a,b) for MNCs by the reference yields the differential scattering cross sections of a MNC along the particular directions ($\theta,\varphi$):

$$\sigma_\perp^{ED}(\theta = \varphi = 90°) = \frac{|S_1(\theta = 90°)|^2}{k^2} = K \frac{A_\perp(\lambda)}{A_\perp^{ref}(\lambda)} \frac{T_{ref}(\lambda)}{T(\lambda)} \sigma_{ref}^{\theta=\varphi=90°}(\lambda) \quad \text{(S2a)}$$

$$\sigma_{//}^{MD+EQ}(\theta = 90°, \varphi = 0) = \frac{|S_2(\theta = 90°)|^2}{k^2} = K \frac{A_{//}(\lambda)}{A_\perp^{ref}(\lambda)} \frac{T_{ref}(\lambda)}{T(\lambda)} \sigma_{ref}^{\theta=\varphi=90°}(\lambda) \quad \text{(S2b)}$$

Superscripts ED and MD+EQ remind that $\sigma_\perp$ and $\sigma_{//}$ are respectively dominated by the scattering of the electric dipole and of the added contribution of the magnetic dipole and electric quadrupole. $K = N_{ref}/N_{MNC}$ is a constant independent of wavelength. The scattering cross section of a silica sphere in water in transverse polarization $\sigma_{ref}^{\theta=\varphi=90°}$ is computed by the Mie theory [36]. The transmission of highly dilute suspensions is dominated by the absorbance of the solvent so that the ratio $T_{ref}(\lambda)/T(\lambda)$ is simply 1. The resulting cross sections (equations (S2a,b)) are plotted in Fig. 3a of the letter.

## 5 – Determination of the dielectric function of silver for T-Matrix numerical simulations of the scattering of a MNC:

The literature offers several data sets for the refractive index of silver [S4]. The most two widely used are found in the Palik handbook compilation [S5,S6] and in the paper by Johnson and Christy [S7]. Unfortunately, they are inconsistent making realistic modeling of the performance of silver in optical applications difficult [S4]. In order to use realistic values of the refractive index in our numerical simulations, we checked the two data sets. We measured the optical extinction of a suspension of silver nanoparticles with a UV-visible spectrometer and we simulated the extinction by the Mie theory. It is important to point out that the silver nanoparticles used in this experiment, synthesized by us, are precisely the ones used to synthesize the MNC satellites (diameter = 25 nm). The Mie theory has no mystery and the simulation requires no particular assumption. The data are shown in Fig. S4. Simulations obtained with the Johnson-Christy data set are far from the experiment and should be discarded. The Palik data set gives a much better agreement and could be used as such. We further improved the dielectric function by introducing a small correction explained in equation (S3) below.

$$\varepsilon_{AgCorr} = \varepsilon_{Palik} + \frac{f_{pPalik}^2}{f^2 + if\gamma_{Palik}} - \frac{f_{Corr}^2}{f^2 + if\gamma_{Corr}} \tag{S3}$$

We first fitted the long wavelength region of the Palik data to a Drude model. A perfect fit is obtained with $f_{pPalik}$ = 2121.15 THz and $\gamma_{Palik}$ = 15.15 THz. We then subtract this Drude component from the permittivity and we add a corrected Drude model with slightly different parameters $f_{Corr}$ = 2131.15 THz and $\gamma_{Corr}$ = 8.15 THz to obtain an almost perfect agreement, as shown in Fig. S5.

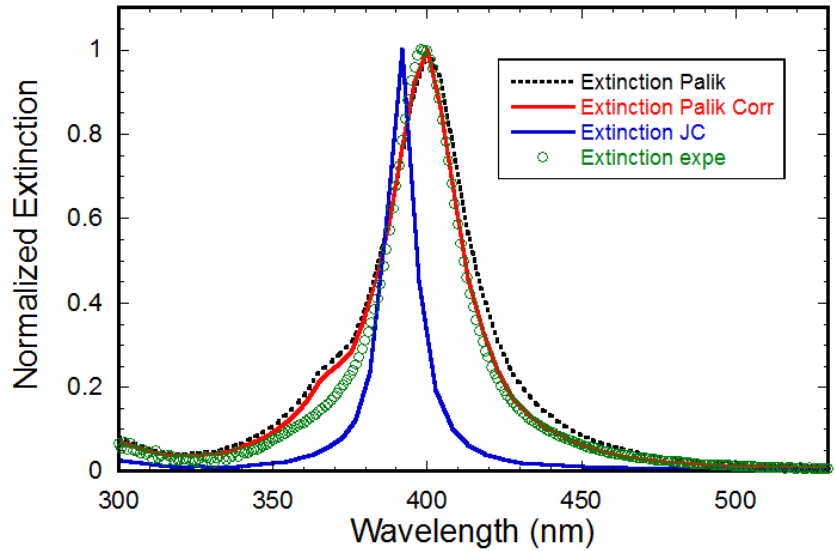

*Figure S5:* experimental (green circles) and simulated extinction of a suspension of silver nanoparticles used in this work. Simulations are performed by Mie theory with Johnson-Christy (blue line) Palik (black dashed line) and Palik-corrected (red line) data sets for the refractive index of silver.

**6 – T-Matrix numerical simulation of the scattering of a single MNC:**

In order to interpret the scattering data, the electromagnetic response of MNCs was computed numerically by a method introduced in reference 24. The MNCs were first constructed by distributing non-intersecting spherical satellites on the surface of the core sphere of diameter 96 nm. Each satellite was constituted of a silver sphere of diameter 25 nm surrounded by a silica shell of thickness 5 nm. The final state is obtained by defining an effective repulsion energy that is subsequently minimized to evenly distribute the satellites by maximizing the distance between them. In a second step, we employed a T-Matrix code developed by Mackowski [35] to simulate the electromagnetic response of the MNCs. It is a FORTRAN code providing the solution to Maxwell's equations for an ensemble of non-overlapping spheres. Following the formalism of the Mie theory describing the scattering of a single sphere [36], the scattered fields are expanded in series of scattering coefficients $a_i^{MNC}$ and $b_i^{MNC}$ computed for the whole cluster. The first order coefficients $a_1^{MNC}$ and $b_1^{MNC}$ correspond respectively to the electric (ED) and magnetic (MD) dipolar mode of the MNC, $a_2^{MNC}$ to the electric quadrupole radiation (EQ) and so on. The refractive indices of silica [S8], water [S9] and ethanol [S10] were taken from the literature. The dielectric function of silver is determined from UV-visible extinction spectra as described is section 4 above.

The computed transverse and axial scattering cross sections of the MNCs $|S_1|^2/k^2$ and $|S_2|^2/k^2$ are displayed in Fig. 3a. The experimental signals are extremely well reproduced for a number of satellites of 22, consistent with TEM images (Fig. 1b).

## 7 – Regression model for ellipsometric data

The ellipsometric measurements were performed in the 240–1000 nm wavelength range with a variable-angle spectroscopic Woollam ellipsometer M2000F. The spot size was set to 50 μm which is small enough to ensure that the footprint of the beam is totally included in the sample of width 100 μm. The semi-infinite geometry enables to derive an exact (complex) expression for the ellipsometric ratio $\rho$ from Fresnel equations:

$$\left(\frac{1-\rho}{1+\rho}\right)^2 \frac{\sin^4\theta}{\cos^2\theta} = \varepsilon\mu\left(\frac{\mu-\varepsilon}{1-\mu\varepsilon}\right)^2 - \sin^2\theta\left(\frac{\mu-\varepsilon}{1-\mu\varepsilon}\right)^2 \tag{S4}$$

in which $\theta$ is the angle of incidence in the ambient air medium of index 1.

For a non magnetic semi-infinite material ($\mu = 1$), expression (S4) can be inverted to yield the well-known formula:

$$N = n + ik = \sqrt{\varepsilon} = \sin\theta_0 \sqrt{1 + \left(\frac{1-\rho}{1+\rho}\right)^2 \tan^2\theta_0} \tag{S5}$$

Hence, measuring the complex ellipsometric reflection ratio $\rho$ at one angle $\theta$ is enough to get the complex refractive index $N = \sqrt{\varepsilon}$. (Normally, $\rho$ is found from the experimentally measured ellipsometric parameters $\Psi$ and $\Delta$ as $\rho = \tan(\Psi)\exp(-i\Delta)$.).

If the medium is magnetic, the problem has one more unknown complex parameter $\mu$ and solving equation (S4) requires measurements of $\rho$ at two angles at least. An analytical formula for permittivity and permeability can then be derived. Let us assume that we have measured the ellipsometric reflection ratios $\rho_1$ and $\rho_2$ at two different angles $\theta_1$ and $\theta_2$. We define

$$A = \left(\frac{\mu-\varepsilon}{1-\mu\varepsilon}\right)^2, B = \varepsilon\mu\left(\frac{\mu-\varepsilon}{1-\mu\varepsilon}\right)^2 \tag{S6}$$

$$X_i = -\sin^2\theta_i, Y_i = \left(\frac{1-\rho_i}{1+\rho_i}\right)^2 \frac{\sin^4\theta_i}{\cos^2\theta_i} \text{ for } i = 1,2. \tag{S7}$$

Equation (S4) yieds:

$$A = \frac{Y_2 - Y_1}{X_2 - X_1}, B = \frac{Y_1 X_2 - X_1 Y_2}{X_2 - X_1} \tag{S8}$$

Inverting equations (S6) is straightforward:

$$\varepsilon = \frac{1}{2}\left(\pm\sqrt{A}\left(1-\frac{B}{A}\right) \pm \sqrt{A\left(1-\frac{B}{A}\right)^2 + \frac{4B}{A}}\right)$$

$$\mu = \frac{B}{A\varepsilon}$$

(S9)

where the signs in (S9) are chosen so as to satisfy the causality principle [S11]. The refractive index can be found as $N = \sqrt{\varepsilon\mu} = \sqrt{B/A}$.

Measuring $\rho$ at more than two angles enables us to increases the accuracy and to test the locality, as explained in the main text. We return to the expression (S4) and plot the linear dependence $Y = AX + B$ with $Y$ and $X$ being given by experimental data measured at different angles. The extracted values of $A$ and $B$ from the linear regression are used in equation (S9) to retrieve $\varepsilon$ and $\mu$.

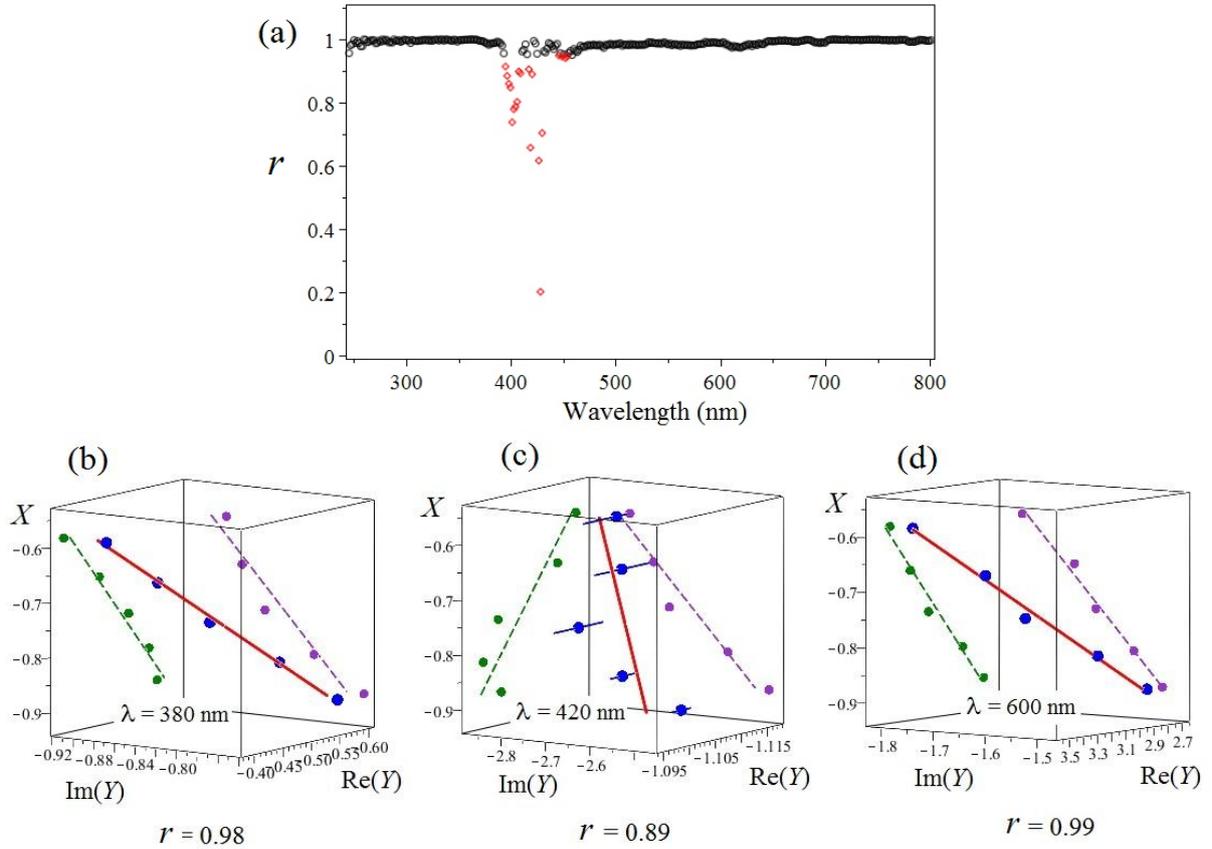

*Figure S6:* (a) Linear Pearson correlation coefficient r showing the goodness of fit across the range of visible wavelengths. Data points with r>0.95 (black circles) correspond to symbols in Fig.2 of the article. Red diamonds denote data points with r<0.95. (b-d) Illustration of the linear relationship Y=AX+B (solid line) at three different wavelengths: (b) λ = 380 nm, r = 0.98 (c) λ = 420 nm, r = 0.89 (poor correlation) and (d) λ = 600 nm, r = 0.99. Symbols are experimental data collected at five angles of incidence. Y, A and B are complex numbers. The projections of the linear regression on the planes (X,Re(Y)) and (X,Im(Y)) are shown (dashed lines).

Fig. S6a shows the spectral evolution of the correlation coefficient of the fit. This coefficient may be viewed as a test of how non-local the retrieved electromagnetic parameters are. Fig. S6b-d illustrates situations of good (b,d) and poor (c) linear correlation.

## 8 – Scattering of the homogeneous sphere (HS):

Here, we consider a fictitious homogeneous sphere (HS) of unknown permittivity $\varepsilon_{HS}$ and permeability $\mu_{HS}$ and determine $\varepsilon_{HS}$ and $\mu_{HS}$ such that the MNC and the effective particle HS exhibit identical scattering. This requires that the coefficients of the scattering matrix of the MNC $S_1^{MNC}$ and $S_2^{MNC}$ computed by the T-Matrix code [35] should be identical to the scattering coefficients of the homogeneous sphere $S_1^{HS}$ and $S_2^{HS}$ given by the usual Mie theory. This condition can be satisfied at first order only (i.e. dipolar) and yields:

$$\frac{3}{2}a_1(\varepsilon_{HS},\mu_{HS}) = S_1^{MNC}(dipole)$$
$$\frac{3}{2}b_1(\varepsilon_{HS},\mu_{HS}) = S_2^{MNC}(dipole)$$
(S10)

where $a_1$ and $b_1$ are the usual Mie scattering coefficients of the homogeneous sphere [36] and $S_i^{MNC}(dipole)$ denotes the truncation at dipolar order of the $S_i$ coefficients in the T-Matrix expansion [35]. For scatterers of radius $R$ much smaller than the wavelength $\lambda$, $a_1$ and $b_1$ are given by a well known series expansion in powers of the Mie parameter $x = 2\pi R N_b/\lambda$ [36], where $N_b$ is the refractive index of the ambient medium. Unfortunately, the condition x<<1 is not fulfilled for MNCs and equation (S10) is solved numerically. The resulting exact numerical solutions $\varepsilon_{HS}(\lambda)$ and $\mu_{HS}(\lambda)$ are displayed on Fig. 3(b,c). Consequently, the ED and MD scattering efficiencies of the MNC and of the HS are rigorously equal (Fig. S6). This is not true for higher order scattering coefficients. In particular, the quadrupolar electric modes (EQ) of the MNC and of the HS are different. Interestingly, equation (S10) can be solved for different sizes, which offers an opportunity to tune the EQ contribution of the HS. We chose a value $R_{HS} = 43$ nm, that pushes the quadrupolar contribution out of the range of experimental wavelengths, (see Fig. S7) while still remaining close to the volume average radius of the MNC ($R_{MNC} = 53.5$ nm).

The numerical simulations of the optical properties of the assembled material shown in Fig. 4 of the letter are hence based on the exact electric and magnetic dipolar polarizations of the MNCs while neglecting the quadrupolar polarization. This may explain some discrepancy observed in the region 400 - 450 nm where the EQ mode is maximal (see inset of Fig. 3a). Fortunately, the ED and MD modes are always dominant and the orders of magnitudes of the simulated parameters are definitely relevant.

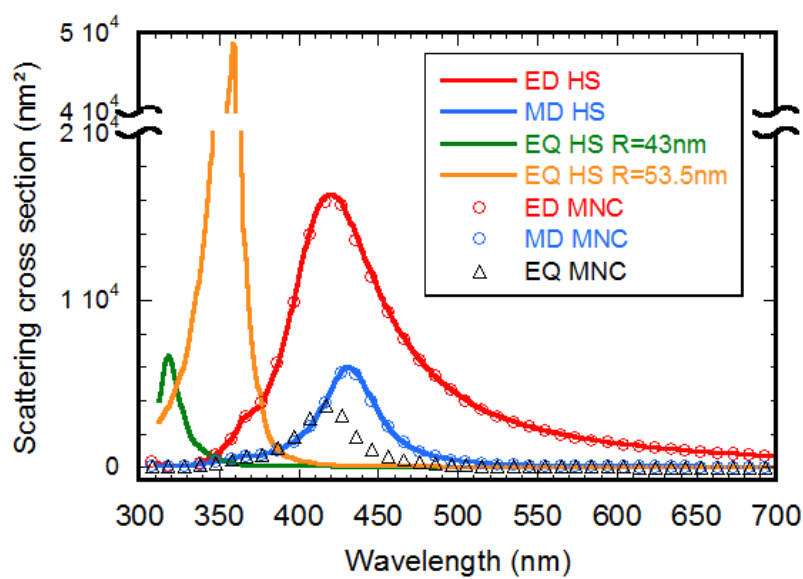

*Figure S7:* *Scattering cross sections computed for the MNC by the T-Matrix code (symbols) and for equivalent homogeneous spheres HS of two different radii R=43 nm and R=53.5 nm by the Mie theory (lines). The electric (ED) and magnetic (MD) dipolar scattering are rigorously identical for the three nano-objects. The electric quadrupolar scattering (EQ) of the MNC (black triangles) is not reproduced by the HS of equivalent radius R=35.5 nm (orange line). A smaller radius (R=43 nm, green line) reduces the EQ scattering of the HS and pushes it towards lower wavelengths so that it does not contribute in the experimental range of wavelengths 380-800 nm.*